\documentstyle[aps,prb,eqsecnum]{revtex}

\tighten

\begin{document}

\draft

\title{Spectral correlations in the crossover between GUE and Poisson 
regularity: on the identification of scales}

\author{Thomas Guhr and Axel M\"uller--Groeling}

\address{Max--Planck--Institut f\"ur Kernphysik, Postfach 103980,
                       D--69029 Heidelberg, Germany}
 
 
\maketitle

\begin{abstract}
Motivated by questions of present interest in nuclear and condensed
matter physics we consider the superposition of a diagonal matrix
with independent random entries and a GUE. The relative strength of
the two contributions is determined by a parameter $\lambda$
suitably defined on the unfolded scale. Using results for the
spectral two--point correlator of this model obtained in the
framework of the supersymmetry method we focus attention on two
different regimes. For $\lambda\ll 1$ the correlations are given by
Dawson's integral while for $\lambda\gg 1$ we derive a novel
analytical formula for the two--point function. In both cases the
energy scales, in units of the mean level spacing, at which
deviations from pure GUE behavior become noticable can be
identified.  We also derive an exact expansion of the local level
density for finite level number.
\end{abstract}

\pacs{PACS numbers: 05.40.+j, 05.45.+b, 72.15.Rn}

\section{Introduction}
\label{sec1}

Transitions from regular to chaotic fluctuation properties have been
observed, experimentally and theoretically, in many different areas of
physics. As examples, we mention topics in atomic (see,
e.g.~Ref.~\onlinecite{Wintgen}) and nuclear (see,
e.g.~Refs.~\onlinecite{Leander,GWP,MDHF}) physics, the relation
between level statistics and localization in condensed matter physics,
and, of course, quantum chaology.  Although the physical mechanisms
responsible for this crossover can differ considerably from system to
system, its statistical signatures follow a common pattern. This fact
calls for a description in terms of Random Matrix
Theory~\cite{Mehta,FH}. Although this transition has already been
addressed by numerous authors over the years, there are still many
formerly unknown features being found in new areas prompting new
interest and research. In particular, there is a constant challenge
for theorists to check whether these new aspects are in agreement or
in disagreement with the predictions of Random Matrix Theory and,
whether or not they can be understood analytically. Because of its
statistical foundations, Random Matrix Theory is a very ambitious
concept since it aims, very often successfully, at a unifying
description of statistical features of very different physical
systems.

Recently, such a challenge has been posed by new developments in
quantum chaology and condensed matter physics. In studying a random
matrix model describing the transition from regular to chaotic
behavior which was motivated by a model of rotational damping in
nuclear physics~\cite{LDB}, Persson and \AA berg~\cite{Aberg1} and
Mizutori and \AA berg~\cite{Aberg2} identified, numerically and
phenomenologically, a critical length scale $L_c$ (there referred to
as $L_{\rm max}$) in the energy spectrum.  Below this scale, the long
range correlations are chaotic and above it they become regular. This
length scale was related to the spreading width or ``energy
localization length'' of the wave functions. The new findings in
condensed matter physics require a more detailed excursion.  In the
last ten years random matrix theory has found wide applications in the
modeling of transport and equilibrium properties of mesoscopic
systems. Prominent examples include electron transport in quasi--one
dimensional disordered wires \cite{q1d} and persistent currents in
mesoscopic rings \cite{pc}.  The general assumption in all these
approaches was that electron--electron interactions can be neglected.
In view of the considerable discrepancy between the relatively small
currents calculated for rings with noninteracting electrons
\cite{pert,pc} and the large experimental values \cite{exp} it has
become generally accepted in the last few years that interactions play
a vital role in the persistent current problem. Quite recently,
evidence has been put forward that interactions might also change our
present understanding of transport and localization in mesoscopic
systems considerably: Shepelyansky predicted \cite{shep} that two
interacting particles in a 1d disordered chain can be extended on a
scale $L_2$ far exceeding the one--particle localization length $L_1$.
Subsequent work \cite{imry,saclay} has quickly led to a definite
confirmation and a more detailed understanding of this phenomenon. As
far as the two--particle effect in one dimension is concerned, there
is probably only one unresolved issue left: the question of the
parameter combination that governs the enhancement of $L_2$ as
compared to $L_1$. Measuring lengths in units of the lattice spacing,
and the strength $U$ of the (Hubbard--type) interaction in units of
the hopping integral, Shepelyansky found $L_2 \propto U^2 L_1^2$. On
the other hand, a microscopic, numerical calculation \cite{opp}
revealed that $L_2 \propto \vert U\vert L_1^2$. In a recent paper
\cite{PW}, an attempt was made to resolve this discrepancy by studying
both an effective Hamiltonian for the two--particle problem and a
microscopic model. The effective Hamiltonian, which was the basis for
Shepelyansky's original claim, is constructed in the following way:
Diagonalize the noninteracting part of the two--electron problem and
express the microscopic Hamiltonian in the basis of two--electron
product states.  The resulting representation consists of a diagonal
contribution containing the eigenvalues of the noninteracting problem,
and an offdiagonal contribution originating from the interaction
operator. With the crucial assumption that both the diagonal and the
offdiagonal matrix elements can be chosen to be random variables we
arrive at a random matrix model that is precisely of the form to be
studied in the present paper. The relative strength between diagonal
and offdiagonal matrix elements is determined by the interaction
strength $U$ and the one--particle localization length $L_1$ (i.e.,
the strength of the disorder). The quantity of interest in this matrix
model is the energy scale $E_c^{(2)}$ at which the spectral
correlations deviate from pure random matrix behavior. This
``spreading width'' or ``two--particle Thouless energy'' was shown by
Imry \cite{imry}, who generalized the famous Thouless scaling picture
to two--electron transport, to determine the two--particle
localization length via the relation
\begin{equation}
\frac{L_2}{L_1} = g_2(L_1) = \frac{E_c^{(2)}}{\Delta_2} \ .
\label{eq1.1}
\end{equation}
Here, $g_2$ is the ``two--particle conductance'' \cite{imry} and
$\Delta_2$ the two--particle level spacing. If one employs Fermi's
Golden Rule to estimate $E_c^{(2)}/\Delta_2 \propto U^2L_1$ (see
Ref.~\onlinecite{imry}), Eq.~(\ref{eq1.1}) immediately leads to
Shepelyansky's original relation $L_2 \propto U^2L_1^2$. The main
result of \cite{PW} was to point out that Fermi's Golden Rule is
inappropriate as long as the off-diagonal matrix elements are
sufficiently weak. Instead, it was found numerically that there exists
a regime where $E_c^{(2)}/\Delta_2 \propto \vert U\vert \sqrt{L_1}$.
While this observation does not directly explain the numerical scaling
law \cite{opp} $L_2 \propto \vert U\vert L_1^2$ it hints at a possible
origin of the linear dependence on $U$. In a recent paper \cite{jacq}
Jacquod, Shepelyansky, and Sushkov argue that $E_c^{(2)}/\Delta_2
\propto \vert U\vert$ around the middle of the spectrum (i.e. in the
vicinity of the band center) while $E_c^{(2)}/\Delta_2 \propto U^2$
otherwise. This offers an alternative explanation for the different
numerical findings. The quantity $E_c^{(2)}/\Delta_2$ should of course
be identified with the critical length scale $L_c$ in
Refs.~\onlinecite{Aberg1,Aberg2}.

It is the purpose of the present paper to derive analytical formulas
that can serve to study the critical length scales discussed in
quantum chaology~\cite{Aberg1,Aberg2} and condensed matter
physics~\cite{PW}. To this end, we make use of an integral
representation of the exact two--point correlation function for the
effective Hamiltonian recently derived by one of us~\cite{TGPRL,TGAP}.
In taking such an analytical approach we wish to present a unifying
discussion which is applicable to the findings in both fields, in
quantum chaology and condensed matter physics. Therefore, since we
want to make this presentation self-contained and readable for a
broader audience, we review and comment on earlier analytical results
in Sec.~\ref{sec2} before we focus on the two different regimes in
Sec.~\ref{sec3}.  In Sec.~\ref{sec4}, we compute the local level
density exactly for finite level number. We summarize and discuss our
results in Sec.~\ref{sec5}.

\section{Model and Outline of Method}
\label{sec2}

In Sec.~\ref{sec2a}, we compare those two random matrix models
describing a crossover from Poisson regular to chaotic fluctuations
which are currently being used in the framework set in the
introduction. We discuss earlier results for the joint probability
density and the correlations in Sec.~\ref{sec2b}. In Sec.~\ref{sec2c},
we briefly review the analytical solution derived in
Refs.~\onlinecite{TGPRL,TGAP}.

\subsection{Two Models - Same Physics}
\label{sec2a}

The Hamilton matrix of our physical problem is represented by an $N
\times N$ matrix $H$ with random entries.  To achieve meaningful
statistical results, we have to average of the ensemble of all those
random matrices which satisfy the physically relevant boundary
conditions. Thus, the physics of such an statistical model is uniquely
determined by choosing the probability density function of the
matrices $H$. Here, we are interested in a crossover or transition
between different regimes implying that we have to go beyond the
commonly used Gaussian Ensembles~\cite{Mehta} which describe fully
chaotic systems.  We will mainly study systems under broken
time-reversal invariance since, first, some presently investigated
systems, e.g.~mesoscopic rings in a magnetic field~\cite{PW}, belong
to this class and, second, recent results~\cite{TGPRL,TGAP} make a
detailed analytical discussion of this situation possible.  For our random
matrix model this means that we have to work with Hermitean matrices
whose distribution reaches the limit of the Gaussian Unitary
Ensemble~\cite{Mehta} for some value of the crossover or transition
parameter. There are two models which are currently being used in this
context.

Weinmann and Pichard~\cite{PW} studied numerically the two-level
correlations induced by the probability density function
\begin{eqnarray}
P^{(A)}(H,\mu) &=& \prod_{n=1}^N \frac{1}{\sqrt{\pi}}
                    \exp\left(-H_{nn}^2\right)
       \prod_{n<m} \sqrt{\frac{1+\mu}{\pi}}
            \exp\left(-(1+\mu){\rm Re}^2H_{nm}\right) \nonumber\\
               & & \qquad \qquad \qquad
       \prod_{n<m} \sqrt{\frac{1+\mu}{\pi}}
            \exp\left(-(1+\mu){\rm Im}^2H_{nm}\right)
\label{ma1}
\end{eqnarray}
where we used our freedom to fix the energy scale by choosing a value
of $1/2$ as the variance of the Gaussian distributions for the
diagonal elements.  The parameter $\mu$ alters the variance of the
distribution for the off-diagonal matrix elements in such a way that
this ensemble interpolates between a pure Gaussian Unitary Ensemble
(GUE) for $\mu=0$ and a Poisson Ensemble for $\mu\to\infty$. In the
latter case, all off-diagonal matrix elements are weighted by
$\delta$-functions. The parameters of this model are related to those
of the physical problem of two interacting particles as follows
\cite{PW}. The matrix dimension $N$ is given by $L_1^d(L_1^d+1)/2$, 
the number of symmetrized two--particle states in a $d$--dimensional
system of size $L_1$. The variance $\sigma^2$ of the diagonal elements
is defined by the bandwidth $B$ of the noninteracting problem,
$\sigma^2 \approx B^2/3$. Normalizing this variance to $1/2$ as we have
done above amounts to measuring all energies in units of
$\sqrt{2/3}B$. In these units the variance of the off-diagonal elements
is given by $(1+\mu)^{-1} = 6(U/B)^2 L_1^{-3d}$.

We introduce a new parameter $\nu=1/\sqrt{1+\mu}$
and write the probability density function~(\ref{ma1}) in a more 
convenient form,
\begin{eqnarray}
P^{(A)}(H,\nu) &=& P^{(1)}(H,\nu) \prod_{n=1}^N p^{(d)}(H_{nn},\nu) 
                                \nonumber\\
P^{(1)}(H,\nu) &=& \frac{2^{N(N-1)/2}}{(\pi\nu^2)^{N^2/2}}
             \exp\left(-\frac{1}{\nu^2}{\rm tr}H^2\right) \ .
\label{ma2}
\end{eqnarray}
The distribution $P^{(1)}(H,\nu)$ is a normalized Gaussian distribution
with variance $\nu^2/2$. The normalized distributions
$p^{(d)}(H_{nn},\nu)$ affect only the diagonal elements. The 
definition~(\ref{ma2}) includes~(\ref{ma1}), but is slightly more general 
since we do not specify the function $p^{(d)}(z,\nu)$ explicitly. 
The Poisson case is now recovered in the limit $\nu\to 0$ whereas
large values of $\nu$ yield the GUE. In the sequel, we will 
refer to the class of models defined by Eq.~(\ref{ma2}) as to model A.
The $k$-point correlation functions depending on $k$ energies 
$x_p, \ p=1,\ldots,k$ in this model are given by
\begin{eqnarray}
R_k^{(A)}(x_1,\ldots,x_k,\nu) = \frac{1}{\pi^k}
             \int d[H] \, P^{(A)}(H,\nu) \,
             \prod_{p=1}^k {\rm Im}\,{\rm tr}\frac{1}{x_p^--H} 
\label{rka}
\end{eqnarray}
where the energies are given imaginary increments such that 
$x_p^- = x_p-i\varepsilon$. The volume element $d[H]$ is, as usual, the
product of all independent variables.

The second model, which we will refer to as model B, is defined
through the random Hamiltonian
\begin{equation}
H(\alpha) \ = \ H^{(0)} \, + \, \alpha H^{(1)}
\label{ha}
\end{equation} 
consisting of a regular and a chaotic part, $H^{(0)}$ and $H^{(1)}$,
respectively, with an interpolating transition parameter $\alpha$.
The matrices $H^{(1)}$ belong to the GUE, i.e.~the probability
density function is given by $P^{(1)}(H,1)$ as defined in Eq.~(\ref{ma2}).
The probability density function $P^{(0)}(H^{(0)})$ is completely
arbitrary. The calculation of the correlation functions in this model,
\begin{equation}
R_k^{(B)}(x_1,\ldots,x_k,\alpha) \ = \
      \frac{1}{\pi^k} \int d[H^{(0)}] \, P^{(0)}(H^{(0)}) 
              \int d[H^{(1)}] \, P^{(1)}(H^{(1)},1) \,
      \prod_{p=1}^k {\rm Im} \, {\rm tr}\frac{1}{x_p^\pm-H(\alpha)} 
\label{rkb}
\end{equation}
requires an average over both matrices. However, due to the rotational 
invariance of the volume elements, the unitary matrices diagonalizing
$H^{(0)}$ can be absorbed implying that, without loss of generality,
$H^{(0)}$ can be chosen to be diagonal. Since we want to have
Poisson statistics as a limiting case, we can write
\begin{equation}
P^{(0)}(H^{(0)}) \ = \ \prod_{n=1}^N p^{(0)}(H_{nn}^{(0)}) \,
                   \prod_{n>m} \delta({\rm Re} H_{nm}^{(0)})
                               \delta({\rm Im} H_{nm}^{(0)})
\label{ped}
\end{equation}
where the function $p^{(0)}(z)$ is smooth but otherwise arbitrary.
We recover the Poisson and the GUE limits for zero and infinite
transition parameter, respectively.
Model B has been studied by many authors, for a review see
Refs.~\onlinecite{TGPRL,TGAP,FKPT,LS,Lenz,PCSF}.

Although it is intuitively obvious that Models A and B must be closely
related, we feel it is worthwhile to work out the precise relationship
between them. Consider model B.  The form~(\ref{rkb}) of the
correlation functions suggests to use the matrices $H=H(\alpha)$
introduced in Eq.~(\ref{ha}) as new integration variables, we find
\begin{equation}
R_k^{(B)}(x_1,\ldots,x_k,\alpha) \ = \
      \frac{1}{\pi^k} \int d[H^{(0)}] \, P^{(0)}(H^{(0)}) 
              \int d[H] \, P^{(1)}(H-H^{(0)},\alpha) \,
      \prod_{p=1}^k {\rm Im} \, {\rm tr}\frac{1}{x_p^\pm-H} \ .
\label{rkbc}
\end{equation}
Since the probability density $P^{(1)}$ is Gaussian, the diagonal
and off-diagonal elements are, as far as the integration over $H^{(0)}$
is concerned, fully decoupled. In other words, the integration over
the purely diagonal matrix $H^{(0)}$ amounts to nothing but a
convolution of each of the functions $p^{(0)}(H_{nn}^{(0)})$ with 
a Gaussian $g(H_{nn}-H_{nn}^{(0)},\alpha)$ which gives a new
probability density function
\begin{equation}
P^{(c)}(H,\alpha) = \prod_{n=1}^N 
          \int_{-\infty}^{+\infty} p^{(0)}(H_{nn}^{(0)})
                    g(H_{nn}-H_{nn}^{(0)},\alpha) dH_{nn}^{(0)}
\label{pec}
\end{equation}
for the matrix elements on the diagonal. Due to the fact that the
total probability density function is a product of $P^{(c)}(H,\alpha)$
and the remaining Gaussians $g({\rm Re}H_{nm},\alpha)$ and 
$g({\rm Im}H_{nm},\alpha)$ for the off-diagonal matrix elements, it is
obviously of the form~(\ref{ma1}) or, equivalently,~(\ref{ma2}).
Hence, we have convinced ourselves that, after a proper re-definition
of the transition parameters, model A and model B describe the same
physical situation. We may therefore confine our discussion to model B
whose analytical properties are, advantageously, already studied in
great detail. Thus, from now on, we drop the upper index $(B)$. Note
that these considerations can easily be generalized to the cases of
real symmetric or quaternion self adjoint matrices.

Finally, we would like to emphasize that there are more models which
describe a transition from Poisson to GUE statistics.  We mention two
recent works, a new formal model by Moshe et al.~\cite{Moshe} and a
study in the context of disordered systems by Altland and
Fuchs~\cite{AF}.  The fact that the two limiting cases, the Poisson
and the GUE statistics, are the same in all these models does not
imply that the interpolation between these limits is necessarily the
same.  The concrete physical situation dictates which model to choose.

\subsection{On the Computation of the Correlation Functions}
\label{sec2b}

A quantity of central interest in every random matrix model is the joint
probability density function $P_E(X,\alpha)$ of the eigenvalues. 
In the case of Hermitean matrices, this function can be worked out
explicitly for our model B. We diagonalize the matrix $H$ 
by a unitary matrix $U$ such that $H=U^\dagger X U$ where 
$X={\rm diag}(x_1,\ldots,x_N)$ is the diagonal matrix of the eigenvalues.
To express the probability density as a function of the latter alone,
we have to do the integrals over $H^{(0)}$ and over $U$,
\begin{equation}
P_E(X,\alpha)d[X] = \int d[H^{(0)}] P^{(0)}(H^{(0)})
                    \int_{U(N)} P^{(1)}(H-H^{(0)},\alpha) d[H] \ .
\label{pea}
\end{equation}
The off-diagonal elements of $H^{(0)}$ are integrated out trivially
leaving us with the diagonal part which we denote by $X^{(0)}$. 
As is well known, the volume element of $H$ is given by 
$d[H]=\Delta_N^2(X)d[X]d\mu(U)$ where $\Delta_N(X)=\prod_{n<m}(x_n-x_m)$ 
is the Vandermonde determinant. Since the function $P^{(1)}$ is
Gaussian, the integral over the unitary group with Haar measure $d\mu(U)$
is just the Harish-Chandra Itzykson Zuber integral~\cite{HCIZ} which
has been used already in Refs.~\onlinecite{Lenz,MP} to compute joint
probability density functions of the eigenvalues. We arrive at
\begin{equation}
P_E(X,\alpha) = \frac{1}{N!} \int d[X^{(0)}] P^{(0)}(X^{(0)})
                {\rm det}[g(x_n-x_m^{(0)},\alpha)]_{n,m=1,\ldots,N}
                \frac{\Delta_N(X)}{\Delta_N(X^{(0)})}
\label{peaiz}
\end{equation}
where $g(z,\alpha)$ is the normalized Gaussian with variance
$\alpha^2/2$ which was introduced above. Note that, due to the 
symmetries of the integrand, this can be written more conveniently as
\begin{equation}
P_E(X,\alpha) = \frac{1}{\sqrt{\pi\alpha^2}^N}
                \int d[X^{(0)}] P^{(0)}(X^{(0)}) 
                \exp\left(-\frac{1}{\alpha^2}
                   {\rm tr}(X-X^{(0)})^2\right)
                \frac{\Delta_N(X)}{\Delta_N(X^{(0)})} \ .
\label{peaizs}
\end{equation}
The Gaussians yield $\delta$-functions for vanishing variance which
implies that the Poisson case is recovered in the limit
$\alpha\to 0$. For very large transition parameter $\alpha\to\infty$,
the GUE joint probability density function has to re-emerge. This
limit is non-trivial but can be checked using the methods developed 
in Ref.~\onlinecite{Lenz}.

The $k$-level correlation function is given as the $N-k$ 
dimensional integral
\begin{equation}
R_k(x_1,\ldots,x_k,\alpha) \ = \
          \int_{-\infty}^{+\infty} dx_{k+1} 
          \cdots \int_{-\infty}^{+\infty} dx_N \, P_E(X,\alpha) 
\label{rkbpe}
\end{equation}
allowing one to use the formula~(\ref{peaizs}) to compute them. This
approach was taken by Lenz~\cite{Lenz} who managed to express the most
interesting two-level correlations $R_2(x_1,x_2,\alpha)$ as a four
dimensional integral. His calculation is exact for every $\alpha$ and
every $N$. Unfortunately, the integrand involves ratios of
$N$-dependent determinants which rendered the evaluation of the
physically interesting limit of infinitely many levels impossible.
Another approach was recently put forward by Pandey~\cite{PCSF} who
chooses Dyson's Brownian motion model as a starting point. He arrives
at a two dimensional integral representation for the two-level
correlation function on the unfolded energy scale, i.e.~in the limit
$N\to\infty$.  Although his derivation is only sketchy, we have no
doubts that his result is correct. Apparently, he avoids the explicit
use of the Harish-Chandra Itzykson Zuber integral by conducting a
highly non-trivial re-summation of certain representation functions of
$U(N)$.  Yet another technique, the graded eigenvalue method, was used
by one of the present authors in Refs.~\onlinecite{TGPRL,TGAP}. It is
a variant of Efetov's famous supersymmetry method~\cite{Efetov} which
relies on the observation that averages over a random potential can be
most efficiently done by mapping the original field theory onto
superspace. In Ref.~\onlinecite{VWZ} it was shown that averages over
random matrix ensembles lead to the same model in superspace. Thus,
supersymmetry can be viewed as something like the ``irreducible
representation'' of Random Matrix Theory. The graded eigenvalue method
allows therefore for a much faster derivation of spectral correlation
functions, even for finite level number, than any other method we are
aware of. Moreover, its results go beyond those of
Refs.~\onlinecite{Lenz,PCSF}. First, higher correlations can be
studied, the graded eigenvalue method yields a $2k$ dimensional
integral representation for the $k$-level correlation functions.
Second, these integrals have a very convenient and rather compact form
even for finite level number $N$. Third, the unfolding, i.e.~the limit
$N\to\infty$ is trivial and, fourth, the two-level correlations on the
unfolded scale become a double integral for completely arbitrary
initial probability density functions $P^{(0)}(H^{(0)})$.

\subsection{Supersymmetry Method}
\label{sec2c}

In order to make our presentation self-contained for those readers
who are familiar with the main concepts of the supersymmetry method,
we will briefly sketch the crucial steps that were taken in 
Refs.~\onlinecite{TGPRL,TGAP}. Readers with little background in
supersymmetry are referred to Refs.~\onlinecite{Efetov,VWZ}, or,
especially regarding the graded eigenvalue method, to Ref.~\onlinecite{TG}.
By including the real parts of the traces of the Green functions
in Eq.~(\ref{rkb}), we define the functions 
$\widehat{R}_k(x_1,\ldots,x_k,\alpha)$ which always allow
the reconstruction of the physically interesting functions
$R_k(x_1,\ldots,x_k,\alpha)$ as linear combinations 
of the functions $\widehat{R}_k(x_1,\ldots,x_k,\alpha)$. 
Advantageously, those can be written as the derivatives
\begin{equation}
\widehat{R}_k(x_1,\ldots,x_k,\alpha) \ = \ \frac{1}{(2\pi)^k} \,  
      \frac{\partial^k}{\prod_{p=1}^k \partial J_p} \,
      Z_k(x+J,\alpha) \Bigg|_{J=0}
\label{drka}\end{equation}
of a normalized generating function $Z_k(x+J,\alpha)$. The energies
and the source variables are ordered in the diagonal matrices 
$x={\rm diag}(x_1,x_1,\ldots,x_k,x_k)$ and 
$J={\rm diag}(-J_1,+J_1,\ldots,-J_k,+J_k)$, respectively.
The physically relevant correlations $R_k(x_1,\ldots,x_k,\alpha)$
are generated by the function $\Im Z_k(x+J,\alpha)$ where the symbol
$\Im$ stands for the proper linear combination. 
By using the supersymmetry method,
the average over the GUE can be performed directly and the generating 
function acquires the form
\begin{eqnarray}
Z_k(x+J,\alpha) &=& \int d[H^{(0)}] \, P_N^{(0)}(H^{(0)})
      \int d[\sigma] \, Q_k(\sigma,\alpha) \nonumber\\
& & \qquad \qquad \qquad {\rm detg}^{-1}\left((x^\pm+J-\sigma)\otimes 1_N - 
                                  1_{2k}\otimes H^{(0)}\right)
\label{grkas}\end{eqnarray}
where $\sigma$ is a $2k\times 2k$ Hermitean supermatrix
and $1_N$ and $1_{2k}$ are the $N \times N$ and the $2k \times 2k$
unit matrices. The function 
\begin{equation}
Q_k(\sigma,\alpha) \ = \ 2^{k(k-1)} \, 
         \exp\left(-\frac{1}{\alpha^2}{\rm trg}\sigma^2\right)
\label{gpdf}\end{equation}
can be viewed as a normalized probability density function for
the supermatrices. It is useful to shift the matrix $x+J$ into
this Gaussian function and then to diagonalize the supermatrix,
$\sigma = u^{-1} su$ by a superunitary matrix $u$
where the $k$ eigenvalues $s_{p1}, \ p=1,\ldots,k$ in the boson boson
and the $k$ eigenvalues $is_{p2}, \ p=1,\ldots,k$ in the fermion fermion 
sector are ordered in the matrix 
$s = {\rm diag}(s_{11},is_{12},\ldots,s_{k1},is_{k2})$. The volume
elements reads $d[\sigma] = B_k^2(s) d[s] d\mu(u)$~\cite{TG}
where $d[s]$ is the
product of the eigenvalue differentials and $d\mu(u)$ is the invariant 
Haar measure of the unitary supergroup. The square root of the Jacobian, 
here referred to as Berezinian, is given by the determinant
$B_k(s) = \det\left[1/(s_{p1}-is_{q2})\right]_{p,q=1,\ldots,k}$.
In these coordinates, the generating function for the entire
transition can be written as
\begin{equation}
Z_k(x+J,\alpha) \ = \ \int Q_k(u^{-1}su-x-J,\alpha) \,
                           Z_k^{(0)}(s) \, B_k(s) d[s] d\mu(u) 
\label{grkasz}\end{equation}
where the function
\begin{equation}
Z_k^{(0)}(x+J) \ = \ \int d[H^{(0)}] \, P_N^{(0)}(H^{(0)}) \, 
           {\rm detg}^{-1}\left((x^\pm+J)\otimes 1_N - 
           1_{2k}\otimes H^{(0)}\right) 
\label{zks}\end{equation}
is the generating function of the correlations of the arbitrary
ensemble defined through $P^{(0)}(H^{(0)})$. Obviously,
the required average over the unitary supergroup
$u$ in Eq.~(\ref{grkasz}) is just the supersymmetric version of the 
Harish-Chandra Itzykson Zuber integral. This observation is the
main ingredient of the graded eigenvalue method.
Collecting everything, and performing the derivatives with respect
to the source variables as in Ref.~\onlinecite{TG}, we find
\begin{equation}
R_k(x_1,\ldots,x_k,\alpha) \ = \
         \frac{(-1)^k}{\pi^k} \, \int G_k(s-x,\alpha) \,
                  \Im Z_k^{(0)}(s) \, B_k(s)d[s]
\label{cfm}\end{equation}
for non-zero $\alpha$. The case $\alpha=0$ is trivial by construction.
Here, the kernel is given by the Gaussian
\begin{equation}
G_k(s-r,\alpha) \ = \ \frac{1}{\sqrt{\pi\alpha^2}^{2k}} \, 
                  \exp\left(-\frac{1}{\alpha^2}{\rm trg}(s-r)^2\right)
\ .
\label{izks}\end{equation}
In order to calculate the generic fluctuations, we have to unfold the
correlation functions for large level number $N$ by removing the
dependence on the level density. We define new energies $\xi_p =
x_p/D,\ p=1,\ldots,k$ where the mean level spacing $D$ is of the order
$1/\sqrt{N}$.  The transition parameter $\alpha$ is defined on the
original energy scale and has therefore to be unfolded, too. The new,
universal transition parameter $\lambda = \alpha/D$ was first
introduced by Pandey~\cite{Pandey}.  The $k$-level correlation
functions on the unfolded scale
$X_k(\xi_1,\ldots,\xi_k,\lambda) = \lim_{N\to\infty} D^k
                    R_k(x_1,\ldots,x_k,\alpha)$
are then generic, i.e.~translation invariant over the spectrum. 
It is useful to unfold the integration variables $s$ in Eq.~(\ref{cfm}) 
by making the rescaling $s \to s/D$. We arrive at
\begin{equation}
X_k(\xi_1,\ldots,\xi_k,\lambda) \ = \ 
\frac{(-1)^k}{\pi^k} \, \int G_k(s-\xi,\lambda) \,
                  \Im z_k^{(0)}(s) \, B_k(s)d[s]
\label{cfum}\end{equation}
for non-zero $\lambda$ where the unfolded generating function of the
arbitrary correlations is given by
$z_k^{(0)}(s) = \lim_{N\to\infty} Z_k^{(0)}(Ds)$. Hence, we have
expressed the unfolded $k$-level correlation function for the transition 
from arbitrary to GUE fluctuations as a $2k$-fold integral.
Note that the limit of infinitely many levels is trivial due to the
application of supersymmetry. The Gaussian kernel remains
unchanged when going from the original to the unfolded energy scale.
It was shown in Ref.~\onlinecite{TGAP} that the generating function
obeys an exact diffusion equation which is, remarkably, the same
on both energy scales, the Gaussian kernel is the Green function of
this diffusion. 

Here, we are mainly interested in the two-level correlations.
Due to translation invariance~\cite{Mehta} on the unfolded scale,
they depend only on the difference $r=\xi_1-\xi_2$ of the
energies. Therefore, it turns out to be useful to make the
change $s_1=s_{11}+s_{21}$, $t_1=s_{11}-s_{21}$, $s_2=s_{12}+s_{22}$
and $t_2=s_{12}-s_{22}$ of the integration variables. Since
$\Im z_2^{(0)}(s)$ depends only on the differences of the eigenvalues
by construction, it cannot depend on $s_1+is_2$. Thus, we write
$\Im z_2^{(0)}(s)=\Im z_2^{(0)}(s_1-is_2,t_1,t_2)$ which, for reasons of
consistency, should be even in each of the differences $t_1$ and
$t_2$.  By using a standard integral theorem of complex analysis, the 
integrals over $s_1$ and $s_2$ can be performed straightforwardly and 
the two-level correlations can thus be cast into the form
\begin{equation}
X_2(r,\lambda) \ = \ \frac{8}{\pi^3\lambda^2} 
       \int\limits_{-\infty}^{+\infty}
       \int\limits_{-\infty}^{+\infty}
       \exp\left(-\frac{1}{2\lambda^2}(t_1^2+t_2^2)\right) 
           \sinh\frac{rt_1}{\lambda^2} 
           \sin\frac{rt_2}{\lambda^2} 
           \frac{t_1t_2}{(t_1^2+t_2^2)^2} \,            
           \Im z_2^{(0)}(0,t_1,t_2) \, dt_1 dt_2
\label{tpfu}\end{equation}
which is exact for every non-zero value of the transition 
parameter $\lambda$.

We emphasize that all results derived so far are correct for arbitrary
initial correlations
$R_k^{(0)}(x_1,\ldots,x_k)$ or $X_k^{(0)}(\xi_1,\ldots,\xi_k)$. 
We now apply them to the case of the Poisson, i.e.~correlation-free initial
spectrum, whose probability density function is given in Eq.~(\ref{ped}).
A straightforward calculation yields
\begin{eqnarray}
Z_k^{(0)}(s) &=& \left(
       1 \, + \, \frac{\pi}{N} \sum_{p=1}^k b_p(s) 
                               \widehat{R}_1^{(0)}(s_{p1})\right)^N
                                          \nonumber\\
      b_p(s) &=& (is_{p2}-s_{p1}) \, 
          \prod_{q \neq p} \frac{is_{q2}-s_{p1}}{s_{q1}^\pm-s_{p1}^\pm} 
\label{gpez}\end{eqnarray}
where $\widehat{R}_1^{(0)}(x)$ is the Stiltjes transform of $R_1^{(0)}(x)$.
We point out that this is exact for all values of $N$.
By choosing $\widehat{R}_1^{(0)}(x)=\widehat{R}_1^{(1)}(x)$ to
evaluate the limit $N\to\infty$, we find 
\begin{equation}
z_k^{(0)}(s) \ = \ \prod_{p=1}^k \exp\left(\mp i\pi b_p(s)\right) \ .
\label{zkpu}\end{equation}
The signs are determined by the choice of the sign
of the imaginary increment in the Green function. 
In the coordinates introduced above, the initial 
condition for the two-level correlations~(\ref{tpfu}) takes the form
\begin{equation}
\Im z_2^{(0)}(0,t_1,t_2) \ = \ \frac{1}{2} {\rm Re}
 \left(\exp\left(-i\pi\frac{t_1^2+t_2^2}{2t_1^-}\right)-1\right) 
\label{izp1}\end{equation}
which still involves an imaginary increment. In 
Refs.~\onlinecite{TGPRL,TGAP}, Eqs.~(\ref{tpfu}) and~(\ref{izp1})
were used to construct a two dimensional integral representation
for $X_2(r,\lambda)$ in terms of Bessel functions. However, for
the discussion to be performed in the next section, the 
form~(\ref{izp1}) is the more convenient one.

\section{Two different regimes}
\label{sec3}

Both the numerical findings \cite{PW} described in the introduction
and the structure of the integral representation (\ref{tpfu}) suggest
to distinguish between the two cases $\lambda\ll 1$ and $\lambda\gg
1$. In these limits the exact but very cumbersome expression
(\ref{tpfu}) can be substantially simplified.
After introducing the proper observable in
Sec.~\ref{sec3a}, we deal with the small $\lambda$ and large $\lambda$
regimes in Secs.~\ref{sec3b} and~\ref{sec3c}, respectively.

\subsection{Spectral long range correlations}
\label{sec3a}

The spectral long range correlations are particularly well suited for
the study of fluctuation properties in systems which undergo
transitions. The level number variance $\Sigma^2(L)$ and the spectral
rigidity $\Delta_3(L)$ describe~\cite{Bohigas}, as functions of the
interval length $L$ on the unfolded scale, the fluctuations of the
integrated, unfolded level density around its smooth value which is
just $L$. Poisson regularity results, for both observables, in a
linear behavior whereas chaotic correlations yield a logarithmic
dependence on $L$. It is intuitively clear and was shown in the
numerical simulations~\cite{GWP,Aberg1,Aberg2,PW} that the admixture
of chaotic features becomes first visible on shorter scales in the
spectrum. In other words, the larger $L$, the stronger have the
chaotic features to be in order to bring the long range correlations
close to the logarithmic behavior. Thus there is a critical interval
length, near which $\Sigma^2(L)$ or $\Delta_3(L)$ ``bend away'' from
the chaotic behavior to show a linear characteristic typical for the
lack of correlations in the Poisson case. This critical length 
$L_{\rm max}$ or $L_c$, respectively, was introduced and studied in
Refs.~\onlinecite{Aberg1,Aberg2,PW} and is also the quantity of
interest in our discussion. In Ref.~\onlinecite{Aberg1}, the critical
length $L_{\rm max}$ was interpreted as the typical spreading width of
the wave functions on the unfolded energy scale.  In the context of
condensed matter physics~\cite{PW}, the critical length $L_c$ was
identified as the Thouless energy in units of the mean level spacing.
Obviously, these two viewpoints describe closely related physical
situations which are formulated in a unifying language in the random
matrix model. 

We now wish to acquire information about the dependence of the critical 
length on the transition parameter $\lambda$ from the
analytical discussion of the previous section. As is well 
known~\cite{Bohigas}, the level number variance is related to the 
two-level correlations by the formula
\begin{eqnarray}
\Sigma^2(L,\lambda) &=& L - 2\int_0^{L} (L-r) Y_2(r,\lambda) dr
                                           \nonumber\\
     X_2(r,\lambda) &=& 1 - Y_2(r,\lambda) 
\label{sxy}
\end{eqnarray}
where $Y_2(r,\lambda)$ is referred to as the two-level cluster function.
The ``bending away'' from the logarithmic behavior is accompanied
by a change of the curvature of $\Sigma^2(L,\lambda)$. Therefore, we may
identify the point at which the curvature changes with the critical 
length. Thus, we simply have to investigate the second order derivative,
\begin{equation}
\frac{d^2}{dL^2}\Sigma^2(L,\lambda) = -2 \, Y_2(L,\lambda) \ ,
\label{sxyd}
\end{equation}
which is, by construction, just the two-level cluster function.
Therefore we can alternatively determine the zeros of $Y_2(L,\lambda)$
or directly investigate the function $\Sigma^2(L,\lambda)$ in order to
identify the scale $L_c$.

\subsection{Small $\lambda$}
\label{sec3b}

The small $\lambda$ regime can be studied by perturbation theory.
This approach has already been taken in Refs.~\onlinecite{FKPT,LS}
with the result
\begin{eqnarray}
X_2(r,\lambda) & \simeq & \frac{r}{\lambda} \, 
                \exp\left(-\frac{r^2}{2\lambda^2}\right) \,
          \int_0^{r/\lambda} \exp\left(\frac{\eta^2}{2}\right) d\eta
                                     \nonumber\\
               & = & \frac{r}{\lambda} \, 
          \int_0^\infty \exp\left(-\frac{\eta^2}{2}\right) 
                        \sin\frac{r\eta}{\lambda} d\eta
\label{appsl}
\end{eqnarray}
which is a universal function of $r/\lambda$. It was checked in
Ref.~\onlinecite{TGAP} that the analytical expressions~(\ref{tpfu}) 
and~(\ref{izp1}) indeed yield the approximation~(\ref{appsl}) for
small $\lambda$. Since $Y_2(r,\lambda)$ is the deviation from
the Poisson behavior which is just unity, we only have to find 
the zero of this function in the variable $r/\lambda$. We find
$r/\lambda \approx 1.3$ implying that the critical length
behaves as
\begin{equation}
L_c \approx 1.3 \, \lambda
\label{lcls}
\end{equation}
for small values of $\lambda$.
Thus, the critical length is indeed linear in $\lambda$
for small values of the transition parameter in agreement with the
numerical investigations in Ref.~\onlinecite{PW}.

\subsection{Large $\lambda$}
\label{sec3c}

To study the large $\lambda$ regime we evaluate the asymptotic expansion
of $X_2(r,\lambda)$ to first order in the inverse transition parameter
by a saddle point approximation.
To bring Eqs.~(\ref{tpfu}) and~(\ref{izp1}) into a form which is 
manifestly suitable for such a procedure we substitute according to
$t_1 = \lambda^2 t_1'$ and $t_2 = \lambda^2 t_2'$.
Omitting the primes we obtain
\begin{eqnarray}
X_2(r,\lambda) &=& \frac{8}{\pi^3\lambda^2} 
\int_{-\infty}^{+\infty} \int_{-\infty}^{+\infty}
dt_1 dt_2 \,\exp\left(-\frac{\lambda^2}{2} (t_1^2+t_2^2)\right) \sinh rt_1
\sin rt_2                \nonumber\\
& & \quad \frac{t_1t_2}{(t_1^2+t_2^2)^2} \Im 
 z_2^{(0)}(0,\lambda^2t_1,\lambda^2t_2) 
\label{eq3.2}
\end{eqnarray}
with the initial condition given by Eq.~(\ref{izp1}).
A direct saddle point approximation still meets with the problem that
there are solutions of the saddle point equations for which 
$t_1^2+t_2^2 = 0$. To avoid the ensuing singularity in (\ref{eq3.2})
we introduce an additional parameter $\beta$ and a function
\begin{eqnarray}
\tilde{X}_2(r,\lambda,\beta) &=& 
\frac{8}{\pi^3\lambda^2} \int_{-\infty}^{+\infty} \int_{-\infty}^{+\infty}
dt_1 dt_2 \, 
\exp\left( -\frac{\lambda^2}{2}\beta (t_1^2+t_2^2) \right)
\sinh rt_1 \sin rt_2  \nonumber\\
& &\quad t_1t_2 \, \Im z_2^{(0)}(0,\lambda^2t_1,\lambda^2t_2) \ ,
\label{eq3.4}
\end{eqnarray}
in terms of which we have 
\begin{equation}
X_2(r,\lambda) =
\frac{\lambda^4}{4} \lim_{\eta\to\infty}
\int_1^\eta d\beta^\prime \int_{\beta^\prime}^\eta d\beta \, 
\tilde{X}_2(r,\lambda,\beta) \ .
\label{eq3.5}
\end{equation}
In the integrand of Eq.~(\ref{eq3.4}), the product $\sinh rt_1 \sin rt_2$
can be replaced by the expression $-i\exp(-rt_1)\exp(-irt_2)$ because 
the additional
contributions vanish upon integration (they make the integrand odd in 
either $t_1$ or $t_2$ or both).  We obtain 
\begin{eqnarray}
\tilde{X}_2(r,\lambda,\beta) &=&
\frac{2}{\pi^3\lambda^2} 
\int_{-\infty}^{+\infty}\int_{-\infty}^{+\infty}
dt_1 dt_2 \, 
\exp\left(-\frac{\lambda^2}{2} \beta (t_1^2+t_2^2)-rt_1-irt_2 \right)
\nonumber\\ 
& &\quad t_1t_2
\left(
\exp\left(-i\pi\lambda^2\frac{t_1^2+t_2^2}{2t_1^-}\right) - 1 + c.c. \right).
\label{eq3.6}
\end{eqnarray}
The contributions originating from the exponential and the constant in
the square brackets, respectively, have to be treated separately.  We
begin with the more complicated term, namely
\begin{eqnarray}
\tilde{X}^{(a)}_2(r,\lambda,\beta) &=& 
\frac{2}{\pi^3\lambda^2} 
\int_{-\infty}^{+\infty}\int_{-\infty}^{+\infty}
dt_1 dt_2 \, 
\exp\left(-{\cal L}(t_1,t_2)\right) \, 
t_1t_2 \ , \nonumber \\
{\cal L}(t_1,t_2) &=& 
\frac{\lambda^2}{2} \beta (t_1^2+t_2^2)+
i\pi\lambda^2\frac{t_1^2+t_2^2}{2t_1^-}
+rt_1+irt_2 \ .
\label{eq3.7}
\end{eqnarray}
To determine the saddle points we can neglect the imaginary increment
for $t_1$ in the second term of ${\cal L}(t_1,t_2)$. The saddle point
equations $\partial_{t_1}{\cal L} = 0$ and $\partial_{t_2}{\cal L} = 0$
read
\begin{eqnarray}
\lambda^2\beta t_1 + i\pi\frac{\lambda^2}{2}
       \left( 1-\frac{t_2^2}{t_1^2} \right) + r &=& 0 \nonumber\\
\lambda^2\beta t_2 + i\pi\lambda^2\frac{t_2}{t_1} + ir &=& 0 \ .
\label{eq3.8}
\end{eqnarray}
With the abbreviations
\begin{equation}
\gamma = i\pi\lambda^2 \qquad {\rm and} \qquad
t = \lambda^2\beta t_2 + ir 
\label{eq3.8a}
\end{equation}
these equations can be cast into the form
\begin{eqnarray}
& & t_1 = -\frac{\gamma t_2}{t} \ , \nonumber\\
& & t^3 + (\gamma^2-2r\gamma)t - 2ir\gamma^2 = 0 \ .
\label{eq3.9}
\end{eqnarray}
The roots of the cubic equation are given by
\begin{eqnarray}
t^{(1)} &=& -i\gamma  \nonumber\\
t^{(2)} &=& i\frac{\gamma}{2}
  \left( 1+\sqrt{1-8r/\gamma} \right) \ , \nonumber\\
t^{(3)} &=& i\frac{\gamma}{2}
  \left( 1-\sqrt{1-8r/\gamma} \right) \ ,
\label{eq3.10}
\end{eqnarray}
in terms of which the solutions $t_{1,2}^{(i)}$ $(i=1,2,3)$ for the
original variables are easily recovered.

Our saddle point approximation proceeds according to the following scheme,
\begin{equation}
\tilde{X}^{(a)}_2(r,\lambda,\beta) \approx
\frac{2}{\pi^3\lambda^2i} \sum_j t_1^{(j)} t_2^{(j)}
\exp\left(-{\cal L}(t_1^{(j)},t_2^{(j)})\right)
\frac{2\pi}{\sqrt{{\rm det} A^{(j)} }} \ ,
\label{eq3.11}
\end{equation}
where
\begin{equation}
A^{(j)} = \left[
  \matrix{
     \frac{\partial^2{\cal L}}{\partial t_1^2} (t_1^{(j)},t_2^{(j)})  &
 \frac{\partial^2{\cal L}}{\partial t_1 \partial t_2} 
       (t_1^{(j)},t_2^{(j)})  \cr
     \frac{\partial^2{\cal L}}{\partial t_2 \partial t_1} 
       (t_1^{(j)},t_2^{(j)})  &
 \frac{\partial^2{\cal L}}{\partial t_2^2 } 
       (t_1^{(j)},t_2^{(j)})  
         }
         \right]
\label{eq3.12}
\end{equation}
is the matrix defining the form of the Gaussian fluctuations around the
saddle point. The approximation (\ref{eq3.11}) is valid for 
$\lambda \gg 1$ and all $r$. Expressing the exponent, the pre-exponential
terms, and det$A$ in (\ref{eq3.11}) in terms of the saddle point values
(\ref{eq3.10}) we get for $j=1,2,3$
\begin{eqnarray}
{\cal L}(t_1^{(j)},t_2^{(j)}) &=& 
-\frac{r}{2\lambda^2\beta (t^{(j)})^2}
(\gamma - it^{(j)})^2(r+it^{(j)}) \ , \nonumber\\
t_1^{(j)} t_2^{(j)} &=&
-\frac{\gamma t^{(j)}}{\lambda^4\beta^2}
\left( 1-i\frac{r}{t^{(j)}} \right)^2 \ , \nonumber\\
{\rm det}A^{(j)} &=& (\lambda^2\beta)^2
\left( 1 - \frac{t^{(j)}}{t^{(j)}-ir}
        \left(1+\frac{(t^{(j)})^2}{\gamma^2} \right)
\right) \ .
\label{eq3.13}
\end{eqnarray}
At this point we notice that $t^{(3)}$ is not a valid saddle point. This
can be most easily seen in the extreme GUE limit ($\lambda\to\infty$, 
$\vert\gamma\vert\to\infty$), where $t^{(3)} \to 2ir$. Obviously, 
det$A^{(3)} < 0$ in this case, i.e. the matrix $A^{(3)}$ is not positive
definite. Therefore, $t^{(3)}$ does not correspond to a maximum of the
integrand and will be discarded henceforth.

Before we proceed we have to investigate the second contribution in 
(\ref{eq3.6}),
\begin{eqnarray}
\tilde{X}^{(b)}_2(r,\lambda,\beta) &=& 
-\frac{2}{\pi^3\lambda^2} 
\int_{-\infty}^{+\infty} \int_{-\infty}^{+\infty}
dt_1 dt_2 \, 
\exp\left(-\tilde{\cal {L}}(t_1,t_2)\right) \,
t_1t_2 \ , \nonumber \\
\tilde{\cal {L}}(t_1,t_2) &=& 
\frac{\lambda^2}{2} \beta (t_1^2+t_2^2)
+rt_1+irt_2 \ .
\label{eq3.14}
\end{eqnarray}
Here, the saddle point equations are trivial and have the single solution
\begin{equation}
\tilde{t}_1 = -\frac{r}{\lambda^2\beta} 
           \qquad {\rm and} \qquad
\tilde{t}_2 = -\frac{ir}{\lambda^2\beta} \ .
\label{eq3.15}
\end{equation}
Therefore, we get in analogy to (\ref{eq3.13})
\begin{eqnarray}
\tilde{\cal {L}}(\tilde{t}_1,\tilde{t}_2) &=& 0 \ , \nonumber\\
\tilde{t}_1 \tilde{t}_2 &=& \frac{ir^2}{(\lambda^2\beta)^2} \ , 
                                                    \nonumber\\
{\rm det}A &=& (\lambda^2\beta)^2 \ .
\label{eq3.16}
\end{eqnarray}
Finally, collecting our results from (\ref{eq3.10}), (\ref{eq3.13}), and
(\ref{eq3.16}) and introducing a new parameter through
\begin{equation}
\frac{r}{\gamma} = -i\frac{r}{\pi\lambda^2} = -i\kappa \ ,
\label{eq3.17}
\end{equation}
we arrive after some algebra at
\begin{eqnarray}
\tilde{X}(r,\lambda,\beta) &=&
\tilde{X}^{(a)}(r,\lambda,\beta) + \tilde{X}^{(b)}(r,\lambda,\beta)
                                      \nonumber\\
&=&
\frac{4}{\lambda^4\beta^3}
\Bigg(
1-\frac{(2i\kappa+1+\rho)(3+\rho)}{8\sqrt{\rho}}
\exp\left(\frac{i\pi r}{4\beta} \frac{(2i\kappa+1+\rho)(3+\rho)^2}
{(1+\rho)^2} \right)  + {\rm c.c.}  \Bigg)
\label{eq3.18}
\end{eqnarray}
with
\begin{equation}
\rho = \sqrt{1+8i\kappa} \ .
\label{eq3.18a}
\end{equation}
The integration over the auxiliary parameter $\beta$ (see (\ref{eq3.5}))
is straightforward and leads to the final result of this section,
\begin{eqnarray}
X_2(r,\lambda) &=& 1 +
 \Bigg( \frac{2}{(\pi r)^2} 
         \frac{(1+\rho)^4}{(2i\kappa+1+\rho)(3+\rho)^3\sqrt{\rho}} \nonumber\\
& & \qquad \qquad \Bigg( \exp\left( i\pi r \frac{(2i\kappa+1+\rho)(3+\rho)^2}
                               {4(1+\rho)^2} \right) \nonumber\\
& & \qquad \qquad \qquad \qquad
       -i\pi r \frac{(2i\kappa+1+\rho)(3+\rho)^2}{4(1+\rho)^2} - 1
\Bigg) + {\rm c.c.} \Bigg)  \ .
\label{eq3.19}
\end{eqnarray}
Although the decisive parameter governing the deviation from the GUE
limit is $\kappa = r/\pi\lambda^2$, closer inspection reveals that the
critical length $L_c$ is still proportional to $\lambda$.  We thank 
Y.~Fyodorov for this insight~\cite{fyod}. In a first order expansion in
$\kappa\ll 1$, one finds
\begin{equation}
X_2(r,\lambda) \approx 1+ \frac{1}{2(\pi r)^2} \left(
e^{-2\pi r\kappa}\cos 2\pi r + 2\pi r\kappa - 1 \right) \ .
\label{add1}
\end{equation}
For $\kappa\to 0$ this reduces to the GUE limit
\begin{equation}
X_2(r,\lambda\to\infty) = 1- \left(\frac{\sin\pi r}{\pi r}\right)^2 \ .
\label{gue}
\end{equation}
For finite but small $\kappa$, Eq.~(\ref{add1}) shows that the
oscillations in $X_2(r,\lambda)$ are exponentially damped on the scale
$\kappa r \propto r^2/\lambda^2$ so that $L_c$ must behave like
$\lambda$. Numerical evaluations~\cite{gmg} of formula (\ref{eq3.19}),
which did not rely on the assumption $\kappa\ll 1$, confirmed this
view. At this point, an important caveat is in order. In Random Matrix
Theory, the unfolding procedure usually involves the crucial
assumption that the transition parameter $\alpha$ scales like the mean
level spacing as the level number $N$ increases. This is formally
necessary to keep the ratio $\lambda=\alpha/D$, i.e.~the transition
parameter on the unfolded energy scale, fixed when the limit
$N\to\infty$ is taken. Physically, this is a well justified assumption
which means that the chaotic admixture, although affecting the
fluctuation properties considerably, does not change the mean level
spacing. The results of Refs.~\onlinecite{TGPRL,TGAP} which are the
starting point of our present considerations were derived under this
assumption. It is not completely straightforward to relate our
parameter $\lambda$ to the transition parameters of the recent studies
in Refs.~\onlinecite{PW,jacq}. Further research along these lines is
under way~\cite{gmg}.

It is interesting to note~\cite{fyod} that under the assumption $r\ll
\lambda\ll \lambda^2$ Eq.~(\ref{add1}) can be further simplified to
give
\begin{equation}
X_2(r,\lambda\gg r) = 1- \left(\frac{\sin\pi r}{\pi r}\right)^2 + 
\frac{1}{\pi^2\lambda^2} \sin^2\pi r \ .
\label{add2}
\end{equation}
Remarkably, this expression coincides exactly with a result derived by
Kravtsov and Mirlin \cite{kravt} for the spectral two--point function of a
quasi--one dimensional disordered wire close to the universal
Wigner--Dyson limit, provided $\lambda$ is identified (via Thouless'
argument) with the conductance. The third term in Eq.~(\ref{add2})
represents the first nonperturbative correction to Wigner--Dyson
statistics. It is intriguing to further explore this analogy.

Finally, we would like to mention that we have tested the quantitative
applicability of Eq.~(\ref{eq3.19}) by means of extensive
numerical simulations~\cite{gmg}. We found that the number variance
$\Sigma^2(L)$ derived from formula~(\ref{eq3.19})
describes the data up to the largest spectral ranges of about
$L\approx 140$ we investigated, as long as $\lambda\gg 1$. We have
therefore confidence in the usefulness and correctness of
Eq.~(\ref{eq3.19}).

\section{Local Level Density}
\label{sec4}

The local density describes the influence of an admixture on the
$m$-th level, say, of a given spectrum. Consider the Hamiltonian
$H(\alpha)=H^{(0)}+\alpha H^{(1)}$ defined in Eq.~(\ref{ha}) with a
diagonal regular part $H^{(0)}$. If we avoid averaging over $H^{(0)}$,
we can study how the density around the $m$-th level of $H^{(0)}$ is
affected as $\alpha$ (and hence the chaotic features) increase.  This
quantity was studied numerically in Refs.~\onlinecite{PW} and
analytically in Refs.~\onlinecite{jac,sigma}. However, this analytical
derivation relies on the saddlepoint approximation and is therefore
only valid for rather large values of $\alpha$, i.e.~when the level
density is sufficiently smooth.  It is the main purpose of this section
to show that a direct and exact evaluation, valid for all values of
$\alpha$, of the local level density is easily possible by the methods
outlined in Sec.~\ref{sec2}, even for finite level numbers. With this
result we can study the local level density in the regime of small
values $\alpha$ which is not accessible to the saddle point
approximation. Apart from this, the calculation which follows is also
of general interest for matrix models. In this context, we mention
related work on Wigner random band matrices~\cite{FCIC}, which also
focuses on the local density of states. 

We express the local level density including its real
part as the derivative
\begin{equation}
\widehat{R}_{1m}(x,\alpha) = \frac{1}{2\pi} 
    \frac{\partial Z_1(x,J,\alpha)}{\partial J_m} \Bigg|_{J=0}
\label{ld1}
\end{equation}
of the generating function
\begin{equation}
Z_1(x,J,\alpha) = \int d[\sigma] \, Q_1(\sigma,\alpha) \, 
                {\rm detg}^{-1}\left((x^--\sigma)\otimes 1_N
                  -1_2\otimes H^{(0)} +\tau\otimes J\right)
\label{ld2}
\end{equation}
depending on a source field $J={\rm diag}(J_1,\ldots,J_N)$ which
resolves the individual levels $n=1,\ldots,N$. We introduced
the matrix $\tau={\rm diag}(-1,+1)$. As usual, we diagonalize
the Hermitean supermatrix, $\sigma=u^{-1}su$, where
$s={\rm diag}(s_1,is_2)$. For the unitary supermatrix, we use the 
explicit parameterization
\begin{eqnarray}
u = \left[ \matrix{ 1+\beta\beta^*/2 & \beta \cr
                 \beta^* & 1+\beta^*\beta/2 } \right] 
\label{ld3}
\end{eqnarray}
in terms of the complex anticommuting angle $\beta$
which allows us to cast the superdeterminant in Eq.~(\ref{ld2})
to first order in $J$ into the form
\begin{equation}
\prod_{n=1}^N\frac{x-is_2-H_{nn}^{(0)}}{x-s_1-H_{nn}^{(0)}} \,
   \sum_{n=1}^N J_n \left(\frac{1+2\beta\beta^*}{x-s_1-H_{nn}^{(0)}}
               +\frac{1+2\beta^*\beta}{x-is_2-H_{nn}^{(0)}} \right) \ .
\label{ld4}
\end{equation}
The derivative with respect to $J_m$ can now be performed easily.
To integrate over the eigenvalues and angles, we have to take
the boundary contributions, often called Efetov Wegner terms, into
account. As shown in Ref.~\onlinecite{TG4}, this can 
be done by applying Rothstein's theory~\cite{Roth} which yields the 
volume element
\begin{equation}
d[\sigma] = \frac{ds_1ds_2}{(s_1-is_2)^2} d\beta d\beta^*
        \left(1-\beta^*\beta (s_1-is_2) 
        \left(\frac{\partial}{\partial s_1}
                  -i\frac{\partial}{\partial s_2}\right) \right)
\label{ld5}
\end{equation}
in which the term containing the derivatives takes care of the
boundary contributions. Collecting everything, we arrive after a
straightforward calculation at the expression
\begin{equation}
\widehat{R}_{1m}(x,\alpha) = \frac{1}{\pi} \,
                             \frac{1}{x^--H_{mm}^{(0)}} \, + \,
              \frac{\partial}{\partial H_{mm}^{(0)}}
                           \widehat{W}(x,\alpha) 
\label{ld6}
\end{equation}
for non-zero $\alpha$. The first part, the Efetov Wegner
term, is simply the unperturbed Green function of the $m$-th level.
The second part can be written as the derivative of the function
\begin{equation}
\widehat{W}(x,\alpha) = \frac{1}{2\pi^2} \
               \int_{-\infty}^{+\infty}\int_{-\infty}^{+\infty}
               \frac{ds_1ds_2}{(s_1-is_2)^2}
               \exp\left(-\frac{1}{\alpha^2}(s_1^2+s_2^2)\right)
       \prod_{n=1}^N\frac{x-is_2-H_{nn}^{(0)}}{x^--s_1-H_{nn}^{(0)}} \ .
\label{ld8}
\end{equation}
Using the identity
\begin{equation}
\prod_{n=1}^N \frac{x-\nu_n}{x-\mu_n} 
         = 1 + \sum_{n=1}^N \frac{\nu_n-\mu_n}{\mu_n-x} 
               \prod_{l \neq n} \frac{\nu_l-\mu_n}{\mu_l-\mu_n}
\label{ld9}
\end{equation}
for $\mu_n=s_1+H_{nn}^{(0)}$ and $\nu_n=is_2+H_{nn}^{(0)}$, we
obtain
\begin{eqnarray}
\widehat{W}(x,\alpha) &=& \frac{1}{2\pi^2} \ \sum_{n=1}^N
               \int_{-\infty}^{+\infty}\int_{-\infty}^{+\infty}
               \frac{ds_1ds_2}{s_1-is_2}
               \exp\left(-\frac{1}{\alpha^2}(s_1^2+s_2^2)\right)
          \frac{b_n(s,H^{(0)})}{x^--s_1-H_{nn}^{(0)}}  
                                       \nonumber \\
b_n(s,H^{(0)}) &=& \prod_{l \neq n} 
   \left( 1 + \frac{is_2-s_1}{H_{ll}^{(0)}-H_{nn}^{(0)}} \right) \ .
\label{ld10}
\end{eqnarray}
This can be worked out further by introducing the permutation
invariant symmetric functions
$c_l(h), l=0,\ldots,N-1$ of a set of variables 
$h_l, \ l=1,\ldots,N-1$ which are defined through
\begin{equation}
\prod_{l=1}^{N-1} \left( 1 + h_l z \right)
         = \sum_{l=0}^{N-1}c_l(h) z^l \ .
\label{ld11}
\end{equation}
In particular we have 
\begin{equation}
c_0(h) = 1, \quad 
c_1(h) = \sum_{l=1}^{N-1} h_l , \quad
c_2(h) = \sum_{l<l^\prime} h_l h_{l^\prime} , \quad \ldots , \quad
c_{N-1}(h) = \prod_{l=1}^{N-1} h_l \ .
\label{ld11a}
\end{equation}
Since $b_n(s,H^{(0)})$ has the structure of the generating function
on the left hand side of Eq.~(\ref{ld11}) if we choose
$h_l^{(n)} = 1/(H_{ll}^{(0)}-H_{nn}^{(0)})$ for $l \neq n$, we
can express the first of Eqs.~(\ref{ld10}) in the form
\begin{eqnarray}
\widehat{W}(x,\alpha) &=& \frac{1}{2\pi^2} \, \sum_{n=1}^N
                            \sum_{l=0}^{N-1} (-1)^l c_l(h^{(n)})
                                       \nonumber \\
& & \qquad \qquad  \int_{-\infty}^{+\infty}\int_{-\infty}^{+\infty}
        ds_1ds_2 \exp\left(-\frac{1}{\alpha^2}(s_1^2+s_2^2)\right)
        \frac{(s_1-is_2)^{l-1}}{x^--s_1-H_{nn}^{(0)}}  \ .
\label{ld12}
\end{eqnarray}
The integral over $s_2$ yields precisely the Hermitean polynomials
\begin{equation}
H_{l-1}(z) = \frac{2^{l-1}}{\sqrt{\pi}} \, 
    \int_{-\infty}^{+\infty} \exp(-\zeta^2) (z-i\zeta)^{l-1} d\zeta
\label{ld13}
\end{equation}
where, in particular, we have 
$H_{-1}(z)=-\sqrt{\pi}\exp(z^2)(1+{\rm erf}(z))$. The remaining
$s_1$ integral can be viewed as the Stiltjes transform
\begin{equation}
\widehat{K}_{l-1}(z,\alpha) = \frac{1}{\pi} 
         \int_{-\infty}^{+\infty} \frac{\exp(-s_1^2/\alpha^2) 
              H_{l-1}(s_1/\alpha)}{z^--s_1}ds_1 \ .
\label{ld14}
\end{equation}
Hence, both integrals can be done and we arrive at
\begin{equation}
\widehat{W}(x,\alpha) = \frac{1}{\sqrt{\pi}} \, \sum_{n=1}^N
                          \sum_{l=0}^{N-1} 
                          \frac{(-1)^l\alpha^l}{2^l} c_l(h^{(n)})
                          \widehat{K}_{l-1}(x-H_{nn}^{(0)},\alpha) \ .
\label{ld15}
\end{equation}
We are mainly interested in the local level density, i.e.~in the
imaginary part $R_{1m}(x,\alpha)={\rm Im}\,\widehat{R}_{1m}(x,\alpha)$,
which is according to Eq.~(\ref{ld6}) given by 
\begin{equation}
R_{1m}(x,\alpha) = \delta(x-H_{mm}^{(0)}) \, + \,
              \frac{\partial}{\partial H_{mm}^{(0)}}
                           W(x,\alpha) 
\label{ld16}
\end{equation}
where $W(x,\alpha)={\rm Im} \, \widehat{W}(x,\alpha)$. Since the
imaginary part of the Stiltjes transform is here just an integration
over a $\delta$ function, we find
\begin{equation}
W(x,\alpha) = \frac{1}{\sqrt{\pi}} \sum_{n=1}^N
              \sum_{l=0}^{N-1} \frac{(-1)^l\alpha^l}{2^l} \, c_l(h^{(n)})
              \, \exp\left(-\frac{(x-H_{nn}^{(0)})^2}{\alpha^2}\right)
              H_{l-1}\left((x-H_{nn}^{(0)})/\alpha\right) 
\label{ld17}
\end{equation}
for an arbitrary matrix $H^{(0)}$. There is a subtle point here to be 
remarked. According to Eq.~(\ref{ld12}), the function
$\widehat{W}(x,\alpha)$ is even in $\alpha$. This, however, is not
immediately obvious from Eq.~(\ref{ld17}). When we introduced the
Hermitean polynomials~(\ref{ld13}), we used $s_2/\alpha$ as the new
integration variable which eventually led to the term $\alpha^l$
in Eq.~(\ref{ld17}). To be mathematically cleaner, we should have
used $s_2/|\alpha|$ which would have produced the term $|\alpha|^l$. 
Hence, we always assume $\alpha > 0$.

The form~(\ref{ld17}) of the function $W(x,\alpha)$ can be viewed
as an expansion in the increasing complexity of the contributions
due to the matrix $H^{(0)}$ which is reflected in the term 
$\alpha^l c_l(h^{(n)})$. To illustrate this, we consider the case
of very small $\alpha$ in which only the $l=0$ term contributes
significantly. Hence we find with Eq.~(\ref{ld11a})
\begin{eqnarray}
W(x,\alpha) &\simeq& \frac{1}{\sqrt{\pi}} \sum_{n=1}^N
              \exp\left(-\frac{(x-H_{nn}^{(0)})^2}{\alpha^2}\right)
              H_{-1}\left((x-H_{nn}^{(0)})/\alpha\right) 
                                         \nonumber\\
            &=& - \sum_{n=1}^N 
     \left(1+{\rm erf}\left((x-H_{nn}^{(0)})/\alpha)\right)\right)
\label{ld18}
\end{eqnarray}
which immediately implies
\begin{equation}
R_{1m}(x,\alpha) \simeq \frac{1}{\sqrt{\pi\alpha^2}} 
         \, \exp\left(-\frac{(x-H_{mm}^{(0)})^2}{\alpha^2}\right) \ .
\label{ld19}
\end{equation}
Thus, for a very small admixture of $H^{(1)}$, the states
of $H^{(0)}$ are smeared out with a Gaussian shape. This can only be
true as long as these Gaussians do not overlap. As the admixture
is increased with $\alpha$, the Gaussian~(\ref{ld19}) of the $m$-th level
slowly starts feeling the influence of the other levels of $H^{(0)}$
and higher and higher orders in $l$ have to be taken into account.
The complexity of this interaction is described through the
more and more entangled structure of the symmetric functions
$c_l(h^{(n)})$ as $l$ increases. We mention in passing that, as is
easily shown, the odd contributions in $l$ can be neglected, 
provided, the $m$-th level lies in the middle of a very long spectrum.
The shape of the local level density changes due to this coupling to 
the other levels of $H^{(0)}$. In Ref.~\onlinecite{jac,PW}, it was shown
numerically that this shape is Lorentzian, even for relatively small
admixture. This is no contradiction to the Gaussian shape~(\ref{ld19})
which only applies in the case of very small $\alpha$. Pictorially
speaking, the coupling to the other levels of $H^{(0)}$ lifts the
tails of this Gaussian which then becomes a Lorentzian. 
In Ref.~\onlinecite{sigma} this Lorentzian shape was analytically
derived in the case of strong admixture.

A further insight into the expansions~(\ref{ld15}) and~(\ref{ld17}) is
provided by the following consideration. Take a harmonic oscillator
with mean level spacing $D^{(0)}$ as initial condition $H^{(0)}$,
i.e.~a picket fence spectrum with equidistant levels. The symmetric
function $c_l(h^{(n)})$ scales now as $1/(D^{(0)})^l$ implying that the
expansion parameter is now given by $\alpha/D^{(0)}$.  This is a
measure of the chaos producing interaction on the scale of the mean
level spacing of the unperturbed spectrum.  Qualitatively, this does
also apply to arbitrary initial conditions $H^{(0)}$.

\section{Summary and Discussion}
\label{sec5}

In this paper we have investigated spectral properties of a particular
class of random Hamiltonians, namely the superposition of a diagonal
matrix with Poisson statistics and a GUE. The motivation for this
investigation came, apart from the obvious relevance for quantum
chaology, from the fact that this class of matrices can serve as an
ensemble of effective Hamiltonians for problems of current interest in
both nuclear and condensed matter physics. After showing the
equivalence of two alternative definitions of this random matrix
ensemble, we analytically discussed two regimes, small
values of the parameter $\lambda$, corresponding to small values of the
electron--electron interaction $U$ in the problem of coherent pair
propagation \cite{shep}, and the regime of large values of $\lambda$.
In the limit $\lambda \gg 1$ we derived a novel explicit expression
for the two point correlation function. In the crossover regime
$\lambda\approx 1$ this expression and the already known perturbative
result for small $\lambda$ differ from each other and do no
longer give a quantitatively correct picture. Comparison with the
exact (albeit difficult to handle) expression~(\ref{tpfu}) however
reveals that the qualitative features of the crossover are still well
described.  This opens the possibility to study analytically and in
detail statistical measures like the number variance $\Sigma^2$.
Such a discussion, furnished with extensive numerical simulations,
will be presented elsewhere~\cite{gmg}.

In Ref.~\onlinecite{Aberg1}, the transition from regular to chaotic
fluctuations was studied using a random matrix model in which the
total level density undergoes a transition from a sharp Gaussian, in
the regular case, to the Wigner semicircle.  With the help of Fermi's
Golden Rule, the critical length $L_{\rm max}$ was estimated as a
function of the model parameters.  The result in
Ref~\onlinecite{Aberg1} cannot be directly related to our discussion
of the critical scale $L_c$. The chaoticity parameter $\Delta$ in
Ref.~\onlinecite{Aberg1} is defined on the original (and not the
unfolded) energy scale, calling for a proper rescalings of the
numerical and the analytical results. In Ref.~\onlinecite{Aberg2} an
additional time dependence was introduced to study energy dissipation.
Depending on the value of the chaoticity parameter $\Delta$ a time
scale $t_1^*$ is identified. It distinguishes between a regime of
anomalous diffusion proportional to $t^2$ for $t<t_1^*$ and normal
diffusion for $t>t_1^*$ of the energy in the eigenbasis at $t=0$. The
precise relation of these interesting findings to the results
presented here has yet to be investigated. It is quite plausible,
however, that the different (chaotic and regular) regimes in the
spectral statistics manifest themselves in this difference in the
energy diffusion.

Finally we have shown how to calculate, for finite $N$, arbitrary
values of $\alpha$, and without approximation, the local density of
states of the matrix ensemble. For extremely small values of $\alpha$
the shape of the local density of states turned out to be Gaussian.
For larger $\alpha$--values, when the levels on the diagonal of the
matrix are no longer strictly isolated, a Lorentzian shape is expected
\cite{PW,jac,sigma}.

We believe to have demonstrated in this paper that many statistical
properties of the ensemble of random matrices studied here are well
under control. This is particularly important in view of recent
developments \cite{PW,Aberg1,Aberg2}, where the ensemble considered in
our work plays a central role.

\section*{Acknowledgment}

We are grateful to S. \AA berg, J.--L. Pichard, H.A. Weidenm\"uller
and D. Weinmann for helpful discussions. We thank the referee for
an important comment on our asymptotic expansion.

\end{document}